# Evaluating Traumatic Brain Injury Outcomes in Road Traffic Crashes

**Azka Rodoshi Oishi**
Department of Public Health
State University of Bangladesh, Dhaka-1461, Bangladesh
Email: azkarodoshi@gmail.com

**Md Jamil Ahsan**
Department of Civil, Environmental and Construction Engineering
University of Central Florida, Orlando, FL 32816, USA
Email: mdjamil.ahsan@ucf.edu

**Azka Sejuti**
Junior Consultant (Paediatrics)
Sreepur Upazilla Health Complex, Gazipur, Bangladesh
Email: azkasejuti@yahoo.com

**B M Tazbiul Hassan Anik**
Department of Civil, Environmental and Construction Engineering
University of Central Florida, Orlando, FL 32816, USA
Email: bmtazbiulhassan.anik@ucf.edu

**ABSTRACT**

Traumatic Brain Injuries (TBIs) resulting from Road Traffic Crashes (RTCs) can have fatal and disabling effects on patients. In this study, we evaluated the TBIs outcomes of patients involved in RTCs and identify key contributing factors affecting these outcomes. Data on 207 patients recorded by physicians at a tertiary hospital in Bangladesh was collected. A random parameters multinomial logit model with heterogeneity in the means was utilized to assess patients' outcomes in three categories: Non-surgical, Surgical, and Fatal. From the random parameters, the study found that male patients (55.48%) are more likely to experience surgical and fatal outcomes. Male motorcycle users have a higher probability of experiencing fatal consequences. Additionally, 60.94% of incidents on rural roads result in surgeries and fatalities, with nighttime crashes on these roads significantly increasing the likelihood of fatal outcomes. Key factors impacting the likelihood of TBIs outcomes include older age, pedestrian involvement, bus and truck crashes, speeding, wet pavements, overtaking, low visibility, and weekday crashes. The study identified two significant interaction variables that increase the probability of fatal outcomes from TBIs: the interactions between low visibility and bus involvement, and between overtaking and wet pavements. While these factors individually had a higher probability of leading to both surgical and fatal outcomes, together these factors increase the risks to fatalities. Overall, our findings provide more detailed insights about the impact of TBIs outcomes resulting from RTCs and emphasize the need to develop more effective measures to improve road safety and patient outcomes.

## INTRODUCTION

Road traffic crashes (RTCs) represent an alarming concern, particularly in low and middle-income countries, where the incidence is rising more rapidly compared to their developed counterparts. Annually, these crashes are responsible for approximately 1.19 million fatalities, and cause non-fatal injuries to an estimated 20 to 50 million individuals (WHO, 2024). RTCs in Bangladesh cause 4,000 deaths and 5,000 injuries each year as per police reports from Roads and Highway Department (RHD) road networks, whereas the World Health Organization (WHO) projects that the annual number of fatalities from RTCs exceeds 25,000 (Acharjee & Ahsan, 2024; ESCAP, 2024). Among the injuries sustained in these crashes, traumatic brain injuries (TBIs) are particularly severe, significantly contributing to both mortality and long-term disability (Hyder et al., 2007).

TBI is defined as damage to the brain caused by an external mechanical force (B. Lee & Newberg, 2005). Each year, TBIs impact over 10 million individuals (Dunne et al., 2020). RTCs are the predominant cause of TBIs, accounting for approximately 60% of all TBI cases (Dunne et al., 2020; Hyder et al., 2007).The prevalence of TBI as a result of RTCs is particularly pronounced in low and middle-income countries (Barki et al., 2023; Hyder et al., 2007). In Bangladesh, many individuals are hospitalized daily due to RTCs, with a significant number suffering from brain injuries, highlighting the severe road safety challenges faced by such countries (Das et al., 2023).

TBI is a critical concern in the healthcare sector, and numerous studies (Baker et al., 2022; Barki et al., 2023; Das et al., 2023; Dunne et al., 2020; Hyder et al., 2007; B. Lee & Newberg, 2005; Majdan et al., 2013) have been conducted to understand its implications better. TBIs resulting from road crashes can lead to fatal outcomes or long-term disabilities if not promptly and properly treated (Barki et al., 2023; Hyder et al., 2007). The impact of TBIs extends far beyond the individual victims, significantly affecting their families and the broader society. The impacts of TBIs include not only the immediate medical costs but also long-term economic losses. These losses stem from a range of factors, including the loss of productivity due to the victim's inability to work, ongoing medical care costs, and the potential need for long-term rehabilitation and support services. Families of TBI victims often face substantial emotional and financial strain, as they may need to provide care and support for an extended period. Moreover, TBIs contribute to the overall economic burden on a country's economy (Ahsan et al., 2021; Das et al., 2023).

Despite the significant impact of TBIs resulting from road crashes, there is an observed gap in their consideration within traffic safety studies. While these injuries are a direct consequence of road crashes, they often do not receive the same level of attention as other aspects of road safety. This oversight can lead to a lack of proper strategies to prevent TBIs and mitigate their impact. Numerous traffic safety studies have been conducted to identify injury severity resulting from traffic crashes and to determine the risk factors contributing to these incidents. Kamruzzaman et al., (2014) analyzed the injury severity caused by road crashes in Dhaka, Bangladesh, employing an ordered probit model and identified the factors which enhanced injury severity. They classified the injury severity as fatal injury, serious injury, or property damage only. Saha et al., (2021) carried out a study in Dhaka, Bangladesh, focusing on the spatial and temporal analysis of frequency of collision and severity of injury in crashes. Their study concentrated on the trends over time and across different areas for crashes involving pedestrians, public transit, and non-traditional modes of transportation. Rahman et al., (1998) undertook a thorough assessment of morbidity and mortality resulting from injuries in Bangladesh, emphasizing the significant impact of injuries on public health within the context of a developing country. Fountas et al., (2020) investigated how different factors influence injury severities in single-vehicle crashes under varying lighting and weather scenarios in Scotland, United Kingdom.

So far, researchers have evaluated the factors affecting injury severity from RTCs, primarily relying on police-reported crash data, and have not specifically focused on TBIs. This approach has been common in both high-income and low and middle-income countries. While police-reported records provide comprehensive information about crashes, they do not offer detailed insights necessary for medical and policy interventions to improve road safety and patient outcomes. In this context, analyzing TBI records of patients involved in RTCs could offer an alternative means to reduce the incidence of TBIs from RTCs. However, to the best of our knowledge, no studies have conducted such analysis, likely due to the limited availability of this data.

In this study, we aim to identify the impact of factors recorded by physicians in Bangladesh on the TBI outcomes of patients experiencing RTCs. The insights gained from this study can inform policymakers to develop targeted interventions and strategies to reduce the incidence of TBIs from RTCs, thereby enhancing road safety and patient outcomes.

## DATA COLLECTION AND DATA PROCESSING

The study was carried out in Shaheed Ziaur Rahman Medical College Hospital (SZMCH) in Bogura, Bangladesh, a tertiary care medical facility, from November 2020 to January 2021, over a span of three months. Our study included a sample of 207 patients who experienced TBIs as a result of RTCs and were admitted through the Department of Casualty.

Both adult and child patients (who are under 18 years old) were included in the study. Patients who sustained TBIs from causes other than RTCs were excluded from the study. Patients who were referred to another hospital for further treatment were also excluded from the study to ensure consistency and accuracy in the data collected from the single medical facility. Additionally, patients who were unresponsive or brought dead, and did not have any relatives or attendees available to provide adequate information about the patients on admission, were excluded from the study. Upon admission, patients were promptly examined by a physician.

Due to the absence of an electronic medical records (EMR) system at the facility, data collection on essential demographic information about the patients, the cause and location of the trauma, risk factor, pre-hospital status and treatment, and details of surgical procedures and outcomes were obtained from the casualty department and neurosurgical ward at the study site. When needed, data were collected from the admission documents of the initial healthcare centers where the patients were first taken. Information on follow up care was gathered throughout the three-month study period from SZMCH.

Once the data collection phase was completed, the recorded information was transferred onto a Microsoft Excel spreadsheet for further analysis. This process involved carefully entering the data into digital format for statistical evaluation and interpretation.

## DESCRIPTIVE STATISTICS

The dependent variable in this study categorizes patient outcomes into three distinct groups: Non-surgical, Surgical, and Fatal (refer to **Table 1**).

**Table 1. Patient outcomes**

| Outcome of TBIs | Description |
|---|---|
| Non-surgical | Patients who receive non-surgical treatment, which involves methods such as, medication, physical therapy, and treatments other than surgery. |
| Surgical | Patients who undergo surgical treatment. |
| Fatal | Patients who died before reaching the hospital and those who passed away after receiving treatment within the study period. |

The age groups are categorized as follows: less than 30 years, 30 to 65 years, and more than 65 years. According to the Local Government Engineering Department (LGED) of Bangladesh, rural roads are mainly village roads. These roads connect local growth centers, local markets, and each other, facilitating movement, connectivity, and economic activities within rural areas (LGED, 2023). Low visibility refers to the reduced ability to see clearly over a distance, typically caused by weather conditions such as heavy rain or fog. All explanatory variables' short descriptions with summary statistics are provided in **Table 2**.

**Table 2. Summary statistics of each variable**

| Variable | Description | Value type | Mean | Std. Dev. | Max. | Min. |
|---|---|---|---|---|---|---|
| Sex | Male | 1: Male 0: Otherwise | 0.625 | 0.49 | 1 | 0 |
| Age | | | | | | |
| <30 years | Less than 30 years old | 1: <30 years 0: Otherwise | 0.550 | 0.481 | 1 | 0 |
| 30 to 65 years | 30 to 65 years old | 1: 30 to 65 years 0: Otherwise | 0.240 | 0.376 | 1 | 0 |
| >65 years | More than 65 years old | 1: >65 years 0: Otherwise | 0.210 | 0.415 | 1 | 0 |
| Rural road | Rural roads defined by LGED, Bangladesh | 1: Rural road 0: Otherwise | 0.375 | 0.485 | 1 | 0 |
| Low visibility | Reduced ability to see clearly over a distance | 1: Low visibility, 0: Otherwise | 0.105 | 0.307 | 1 | 0 |
| Speeding | faster than the legally established speed limit of the road | 1: Speeding 0: Otherwise | 0.090 | 0.287 | 1 | 0 |

| Variable | Description | Value type | Mean | Std. Dev. | Max. | Min. |
|----------|-------------|------------|------|-----------|------|------|
| Weekday | Days other than Friday, Saturday and any holiday | 1: Weekday 0: Otherwise | 0.660 | 0.347 | 1 | 0 |
| Wet pavement | Road surface that is covered or saturated with water | 1: Wet pavement 0: Otherwise | 0.305 | 0.462 | 1 | 0 |
| Overtaking | Passing another vehicle traveling in the same direction | 1: Overtaking 0: Otherwise | 0.275 | 0.364 | 1 | 0 |
| Bus driver/ passenger | Bus driver/ passenger | 1: Bus 0: Otherwise | 0.285 | 0.452 | 1 | 0 |
| Car driver/ passenger | Car driver/ passenger | 1: Car 0: Otherwise | 0.225 | 0.418 | 1 | 0 |
| Truck driver/ assistant | Truck driver/ assistant | 1: Truck 0: Otherwise | 0.175 | 0.387 | 1 | 0 |
| Motorcycle | Motorcycle rider | 1: Motorcycle 0: Otherwise | 0.210 | 0.408 | 1 | 0 |
| Pedestrian | Pedestrian involvement | 1: Pedestrian 0: Otherwise | 0.105 | 0.307 | 1 | 0 |
| Nighttime | Nighttime | 1: Nighttime 0: Otherwise | 0.580 | 0.495 | 1 | 0 |

## METHODOLOGY

For the methodology, this study considered a random parameters multinomial logit model to estimate the patient's outcome from TBIs. This approach was selected for its ability to account for potential heterogeneity in both the mean and variance of the parameters. However, upon a thorough investigation, no significant variance was detected in the data. The analysis was guided by patient outcomes into three distinct groups in the **Table 1**: Non-surgical, Surgical, and Fatal. By enabling the mean and variance of random parameters to be functions of the independent variables, this estimation method adds flexibility and precision, facilitating a deeper exploration of any latent heterogeneity within the dataset (Ahsan, Abdel-Aty, & Abdelrahman, 2024). Notably, this approach has seen extensive use in related disciplines, such as travel behavior and transportation safety research, as it provides significant insights that fixed-parameter models may overlook, offering a richer understanding of unobserved variability (Barbour et al., 2021; Hossain et al., 2024; J. Lee et al., 2021).

To identify the model that best fits the data and the dependent variable, a function was established to determine the probability of TBI outcomes falling into one of three categories: Non-surgical, Surgical, or Fatal, as shown in Equation (1).

$$F_{kn} = \beta_k \mathbf{X}_{kn} + \varepsilon_{kn} \tag{1}$$

Here, $X_{kn}$ represents a vector of explanatory (independent) variables influencing the likelihood that observation $n$ corresponds to a specific TBI outcome $k$. The term $\beta_k$ denotes the vector of estimable parameters associated with outcome $k$, and $\varepsilon_{kn}$ signifies the disturbance term. Assuming the disturbance term follows a generalized extreme value distribution, a logit model (Eq. 2) can then be defined (Ahsan, Abdel-Aty, & Abdelrahman, 2024; Washington et al., 2020).

$$P_n(k) = \frac{EXP[\beta_k \mathbf{X}_{kn}]}{\sum_{\forall K} EXP[\beta_k \mathbf{X}_{kn}]} \tag{2}$$

where $P_n(k)$ is the probability of patient's TBIs $n$ having an outcome level $k$ (Non-surgical, Surgical, and Fatal).

To address instances where one or more parameter estimates in the vector $\beta$ vary across observations due to unobserved heterogeneity, a distribution for these parameters can be assumed. This allows for capturing variations that are not directly observed in the data. Under this framework, Equation (3) can be expressed as follow (Washington et al., 2020).

$$P_n(k) = \int \frac{EXP(\beta_k \mathbf{X}_{kn})}{\sum_{\forall K} EXP(\beta_k \mathbf{X}_{kn})} f(\beta_k / \varphi_k) d\beta_k \tag{3}$$

In this context, $f(\beta k|\varphi k)$ represents the density function of $\beta_k$, where $\varphi_k$ is the vector of parameters characterizing the mixing density function, specifically capturing the mean and variance. This density function allows for the random parameters $\beta_k$ to vary across observations according to the distribution specified by $\varphi_k$. The inclusion of this density function in the model enables us to account for unobserved heterogeneity by letting the parameter values change across observations, thereby improving the model's flexibility and fit to the data.

To enhance accuracy and add flexibility in capturing unobserved heterogeneity, the $\beta_{kn}$ vector can be modeled (Eq. 4) as a function of variables that influence both its mean and variance. By doing so, $\beta_{kn}$ becomes adaptable to variations across observations, allowing the model to account for differences that might otherwise remain hidden. Specifically, this approach enables the mean and variance of $\beta_{kn}$ to be responsive to certain explanatory variables, leading to more understanding of the effects within the dataset and offering a more precise estimation of TBI outcome probabilities across diverse patient profiles (Barbour & Mannering, 2023; Mannering et al., 2016).

$$\beta_{kn} = \beta_k + \Theta_{kn} \mathbf{Z}_{kn} + \sigma_{kn} EXP(\Psi_{kn} \mathbf{W}_{kn}) \nu_{kn} \tag{4}$$

In this formulation, $\beta_k$ represents the mean parameter estimate across all possible TBI outcome alternatives. The vector $Z_{kn}$ contains observation-specific explanatory variables that introduce heterogeneity in the mean for the safety outcome $k$, with $\Theta_{kn}$ serving as the corresponding vector of estimable parameters. Similarly, $W_{kn}$ is a vector of observation-specific explanatory variables that captures heterogeneity in the standard deviation (or variance) $\sigma_{kn}$ associated with the parameter vector $\Psi_{kn}$. Finally, $v_{kn}$ denotes the disturbance term. This setup allows the mean and variance of the random parameters to be functions of specific explanatory variables, providing additional flexibility in the model by accounting for both observed and unobserved heterogeneity in TBI outcome predictions (Ahsan, Abdel-Aty, & Abdelrahman, 2024; Barbour & Mannering, 2023).

The model estimation in this study employed a simulated maximum likelihood approach with 1,000 Halton draws. To provide deeper insights, marginal effects were calculated for each case and averaged across observations, yielding a comprehensive view of variable impacts and interactions. Likelihood ratio tests confirmed the model's statistical superiority. A normal distribution was assumed for the random parameters, as alternatives such as log-normal, exponential, and uniform distributions did not yield improved results. The normal distribution is also commonly used in random parameter models, reinforcing its selection here.

**RESULTS AND DISCUSSION**

Several statistically significant variables were found in the model. **Table 3** presents the results of the random parameters multinomial logit model with heterogeneity in the means of the random parameters on the patient's outcome. As described in the methodology, three risk categories are evaluated namely: Non-surgical [S], Surgical[M], and Fatal [MM]

**Table 3.** Random parameters multinomial logit model with heterogeneity in the means of random parameters on the patient's outcome: Non-surgical [S], Surgical[M], and Fatal [MM]

| | Estimated parameter | t-Statistics | Marginal effects | | |
|---|---|---|---|---|---|
| | | | Non-surgical [S] | Surgical [M] | Fatal [MM] |
| Constant [S] | 1.754* | 4.07 | | | |
| Constant [MM] | 3.682* | 3.18 | | | |
| **Random parameters in utility functions** | | | | | |
| Male (1 if male, 0 otherwise) [MM] | 0.268* | 3.64 | -0.041 | 0.018 | 0.023 |
| Standard deviation of parameter distribution | 1.946* | 2.57 | | | |
| Rural road (1 if rural, 0 otherwise) [MM] | 0.297* | 3.61 | -0.091 | 0.041 | 0.05 |

| | Estimated parameter | t-Statistics | Marginal effects | | |
|---|---|---|---|---|---|
| | | | Non-surgical [S] | Surgical [M] | Fatal [MM] |
| Standard deviation of parameter distribution | 1.069* | 2.87 | | | |
| **Heterogeneity in the mean of the random parameter** | | | | | |
| Male: Motorcycle (1 if motorcycle involvement, 0 otherwise) [MM] | 1.013* | 4.23 | | | |
| Rural road: Nighttime (1 if nighttime, 0 otherwise) [MM] | 0.735* | 2.62 | | | |
| **Fixed parameters in utility functions** | | | | | |
| *Car driver/passenger Base* | | | | | |
| **Pedestrian** (1 if pedestrian, 0 otherwise) [S] | -0.243* | 4.05 | -0.035 | -0.019 | 0.054 |
| **Bus driver/ passenger** (1 if bus involvement, 0 otherwise) [MM] | 0.018* | 2.83 | -0.047 | 0.011 | 0.036 |
| **Truck driver/ assistant** (1 if truck involvement, 0 otherwise) [MM] | 0.548* | 3.06 | -0.088 | 0.016 | 0.072 |
| **Age: >65 years** (1 if age >65, 0 otherwise) [MM] | 0.173* | 2.80 | -0.061 | -0.016 | 0.077 |
| **Weekday** (1 if weekday, 0 otherwise) [S] | 1.034* | 3.15 | 0.053 | -0.024 | -0.029 |
| **Wet pavement** (1 if wet, 0 otherwise) [MM] | 0.915* | 2.79 | -0.077 | 0.014 | 0.063 |
| **Low visibility** (1 if low, 0 otherwise) [S] | -0.193* | 3.68 | -0.039 | 0.025 | 0.014 |
| **Speeding** (1 if speeding, 0 otherwise) [MM] | 0.507* | 2.69 | -0.028 | -0.012 | 0.04 |

| | Estimated parameter | t-Statistics | Marginal effects | | |
|---|---|---|---|---|---|
| | | | Non-surgical [S] | Surgical [M] | Fatal [MM] |
| **Overtaking** (1 if overtaking, 0 otherwise) [MM] | 1.137* | 3.60 | -0.10 | 0.084 | 0.016 |
| **(Low visibility * Bus)** interaction [S] | -2.06* | 3.97 | -0.065 | -0.012 | 0.077 |
| **(Overtaking * Wet pavements)** interaction [MM] | 0.992* | 4.30 | -0.086 | -0.020 | 0.106 |
| Number of observations | 220 | | | | |
| Log likelihood at zero, LL (0) | -429.86 | | | | |
| Log likelihood at convergence LL(β) | -341.52 | | | | |
| $\rho^2 = 1 - LL(\beta)/LL(0)$ | 0.21 | | | | |
| * 5% level of significance | | | | | |

## Random parameters and heterogeneity in the mean

Considering statistically significant random parameters, the male indicator variable was found to vary across observations, as indicated by the statistically significant random parameter shown in **Table 3**. This random parameter follows a normal distribution and is estimated with a mean of 0.268 and a standard deviation of 1.946, showing that 44.52% of this distribution is below zero and 55.48% is above zero, as determined from the Z-value and Z-table (Hossain et al., 2024). This implies that a significant portion of male patients, approximately 55.48%, are more likely to experience surgical and fatal outcomes. In contrast, the remaining 44.52% have a probability of not requiring surgery. The marginal effect analysis further supports this observation, indicating a higher probability of both surgery and fatal outcomes for male patients (see **Figure 1**). Moreover, the study also highlights that, male individuals, who are motorcycle users, exhibit a higher probability of experiencing fatal consequences. In Bangladesh, motorcycles are a popular mode of transportation and there is a higher prevalence of motorcycle use among males compared to females (Miah et al., 2024). This reveals the increased risk of TBIs outcome associated with motorcycle use among male patients. Several factors contribute to the increased risk faced by male motorcycle users. One significant factor is the low rate of helmet usage. Helmets are crucial for protecting riders from TBIs. Additionally, using mobile phones while driving is a major distraction that can lead to loss of control and collisions, resulting in TBIs.

Another variable 'rural road' indicator variable was found to vary across observations. This random parameter follows a normal distribution and is estimated with a mean of 0.297 and a standard deviation of 1.069, showing that 39.06% of this distribution is below zero and 60.94% is above zero, as determined from the Z-value and Z-table. This implies that for patients involved in crashes, approximately 60.94% of rural road incidents are more likely to result in fatalities from TBIs, while the remaining 39.06% have a probability of not needing surgery. The marginal effect (refer to **Table 3** and **Figure 1**) further supports a higher probability of both surgery and fatal outcomes of TBIs from crashes occurring on rural roads. Additionally, the study emphasizes that nighttime crashes on rural roads have a significantly higher probability of resulting in fatal outcomes from TBIs. This is evidenced by the positive heterogeneity value (1.013) in the mean. Contributing factors might include poorly maintained and narrow roads, lack of essential infrastructure like pedestrian signals and dedicated lanes, inadequate lighting, and insufficient reflective signs. Additionally, limited access to emergency medical services in rural areas often turns survivable TBIs into fatal injuries due to delayed medical attention.

**Fixed parameters without interaction**

*Pedestrian*

The analysis indicates that presence of pedestrian significantly impacted surgical and fatal probabilities. Higher pedestrian activity is positively associated with TBIs outcomes, meaning that as the number of pedestrians increases, the chances of TBIs resulting in the need for surgery or leading to fatalities also increase (Ahsan, Abdel-Aty, & Anwari, 2024). Pedestrians are particularly vulnerable to severe and fatal injuries compared to other road users. This might be due to the lack of the protective barriers provided by a vehicle, making them more exposed to direct impacts in the event of a collision.

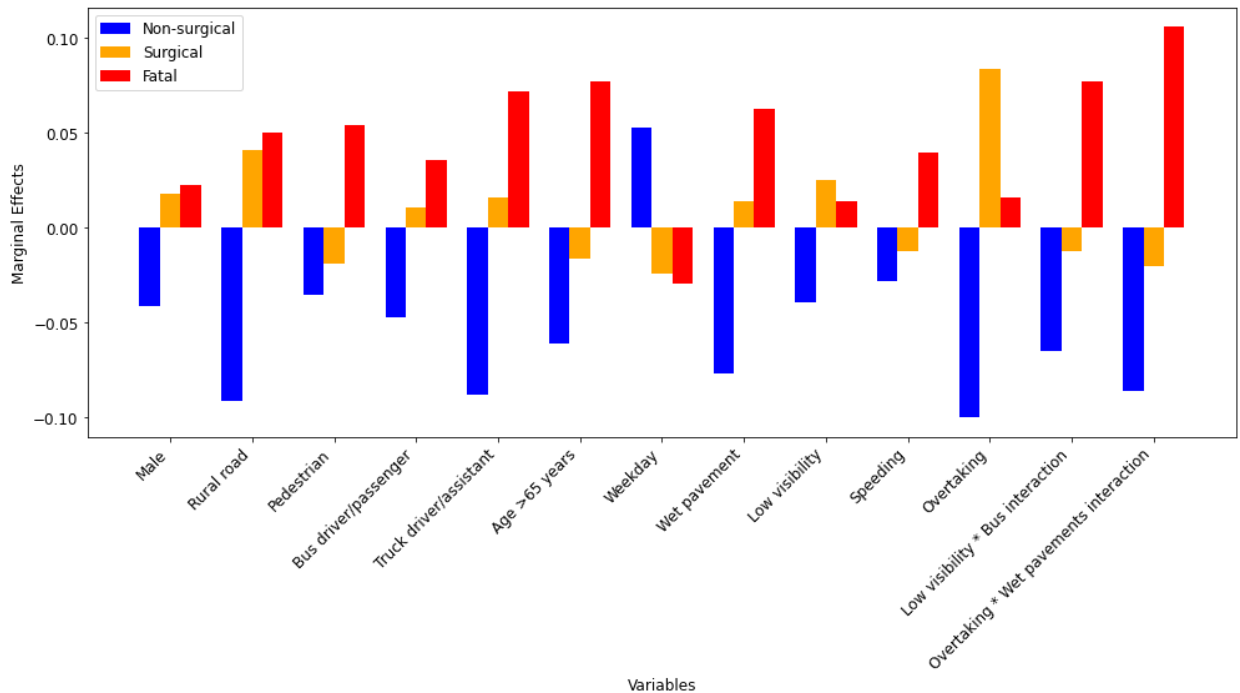

**Figure 1. Marginal effects of the variables used in the model**

### Bus driver/passenger

The analysis indicates that involvement as a bus driver or passenger increases the probability of TBIs that require surgical intervention and the risk of fatal outcomes. Several factors contribute to the safety issues associated with bus transportation in Bangladesh. The buses are often overcrowded, poorly maintained, and lack essential safety features such as seat belts and proper restraints. Speeding is a frequent issue that reduces the driver's reaction time and increases collision severity, leading to more serious traumatic brain injuries and fatalities. Tailgating, or following other vehicles too closely, is another risky behavior commonly observed among bus drivers that makes it difficult to avoid collisions if the vehicle in front suddenly stops or slows down, often resulting in TBIs.

### Truck driver/ assistant

This model demonstrates that involvement with trucks, whether as a driver or an assistant is associated with a higher likelihood of TBIs that require surgical intervention and a greater risk of fatal outcomes (refer to **Table 3** and **Figure 1**). Several factors might contribute to this risk. Trucks, being large and heavy vehicles, generate substantial force upon impact during crashes. This force can cause severe injuries, especially TBIs, to both truck occupants and other road users involved in the collision. Trucks often travel long distances on highways and rural roads, where the chances of TBIs from high-speed crashes are greater. The lack of proper safety features in trucks, such as seat belts and air brakes, contributes to higher surgical and fatality rates of TBIs.

### Age

Our result reveals that the probability of fatality from TBI increases with age (>65 years) which is similar to other studies (Mcintyre et al., 2013; Røe et al., 2013; Skaansar et al., 2020). This elevated risk can be attributed to several factors related to the health and medical condition of older individuals. Older adults often experience a slower recovery process and are more likely to suffer complications following a TBI. Additionally, pre-existing health conditions such as hypertension, diabetes, and cardiovascular diseases can exacerbate the outcomes of a TBI, making it more challenging for the body to heal and increasing the risk of fatality. Moreover, older individuals may not be fit for high-intensity treatments or surgeries required to address TBIs.

### Weekday

The analysis reveals that road crashes causing TBI on weekdays in Bangladesh significantly lower the probabilities of surgical intervention and fatal outcomes. On weekdays, especially during peak hours, traffic congestion is higher, which generally reduces vehicle speeds. Lower speeds can lead to less severe crashes, resulting in TBIs that are less likely to require surgery or result in fatalities. Weekdays typically have better availability of emergency services and medical staff, as hospitals and clinics are fully operational. Prompt medical response and immediate access to medical facilities can lead to quicker treatment, reducing the impact of TBIs and the need for surgical interventions.

### Wet pavement

Our result indicates that the presence of wet pavement significantly increases the surgical and fatal probabilities from TBIs in a crash. Wet pavement decreases the friction between tires and the road surface, leading to reduced traction. This makes it more difficult for drivers to control their vehicles, increasing the likelihood of skidding and losing control, which can result in brain injury. Vehicles require longer distances to come to a complete stop on wet roads. This can lead to rear-end collisions and other high-impact crashes, which are more likely to cause TBIs and fatal injuries.

### *Low visibility*

The findings indicate that low visibility significantly increases the surgical and fatal probabilities of TBIs resulting from crashes. Low visibility conditions, such as fog, heavy rain, or nighttime darkness, reduce the distance drivers can see ahead. This decreases the time available for drivers to react to obstacles, other vehicles, or sudden changes in road conditions, leading to a higher likelihood of severe crashes. In low visibility conditions, drivers may not notice hazards until it is too late to avoid them. This often results in high-speed collisions or sudden braking, which can cause severe impacts and lead to traumatic brain injuries. The force of these collisions increases the probability of fatal outcomes.

### *Speeding*

The analysis shows that speeding substantially raises the likelihood of fatal TBIs resulting from crashes. Speeding amplifies the force of impact in collisions. This increased force is particularly damaging to the brain, resulting in a higher likelihood of TBIs and fatal outcomes. At higher speeds, vehicles are more difficult to control. This loss of control can lead to more frequent and severe crashes, including rollovers and side-impact collisions, which are particularly dangerous and more likely to result in traumatic brain injuries. Many roads in Bangladesh are not designed for high-speed travel. Speeding on poorly maintained roads, which may have potholes, sharp turns, and inadequate signage, increases the risk of crashes. These hazardous road conditions, combined with speeding, can lead to TBIs with fatal outcomes.

### *Overtaking*

The analysis indicates that overtaking significantly increases the risk of surgical and TBIs in crashes. Overtaking often involves crossing into oncoming traffic lanes, which significantly raises the risk of head-on collisions. These types of collisions are particularly severe due to the combined speed of both vehicles, leading to TBIs. When overtaking, drivers may have limited visibility of the road ahead, that can prevent drivers from seeing oncoming traffic or obstacles in time to avoid a collision, increasing the chances of high-impact crashes that cause TBIs. Narrow roads and lack of passing lanes can contribute to the dangers of overtaking and increase the likelihood of fatalities.

### **Fixed parameters with interaction effect**

This analysis identified two significant interaction variables that impact the probability fatal incidents resulting from TBIs. The marginal effects shown in **Figure 1** illustrate the following findings:

***Bus Involvement and Low Visibility*:** Individually, both bus involvement and low visibility conditions significantly increase the likelihood of surgical and fatal incidents. When these two factors combine, resulting in crashes involving buses under low visibility conditions, the probability of fatal incidents from TBIs is considerably increased.

***Overtaking and Wet Pavements***: Individually, both overtaking and wet pavements significantly raise the chances of surgical and fatal incidents due to the increased risk of losing control and colliding with other vehicles. When these factors occur together by overtaking on wet pavements, the risk of fatal incidents from TBIs is markedly raised.

## CONCLUSIONS AND FUTURE WORK

RTCs, particularly in low and middle-income countries, significantly contribute to TBIs, posing a major public health concern. In the US, TBIs cause 1.35 million emergency visits, 275,000 hospitalizations, and 52,000 deaths annually, making up 30.5% of all TBI related deaths and significantly increasing healthcare costs (Soares De Souza et al., 2015). In low and middle-income countries, which comprise 85% of the world's population, TBIs account for 90% of global deaths from these injuries, making the costs significantly higher (de Silva et al., 2009). Therefore, this burden is greater in Bangladesh compared to high-income countries. Despite numerous studies on traffic safety, there is a critical gap in addressing TBIs outcomes within this context. This study aims to identify the contributing factors to TBIs outcomes resulting from road crashes and to draw the attention of policymakers to this issue. By integrating these findings into traffic safety research and policymaking, the goal is to create more effective measures to reduce the incidence and impact of TBIs, ultimately enhancing public health and safety.

For this study, data were collected from a tertiary medical college hospital, ensuring that only observations with complete information necessary for the analysis were processed. A random parameters multinomial logit model with heterogeneity in the means was employed to assess patient outcomes following traffic crashes. The outcomes were categorized into three risk levels: Non-surgical, for patients treated with medication, physical therapy, or other non-surgical methods; Surgical, for patients who underwent surgical procedures; and Fatal, for patients who died either before reaching the hospital or after receiving treatment during the study period.

The analysis revealed several key factors impacting the likelihood of surgical and fatal TBIs resulting from traffic crashes. The presence of pedestrians, bus and truck involvement significantly increase the risk of TBIs outcomes. Age is another important factor, with older adults facing higher fatality rates. On weekdays, traffic crashes typically result in less critical TBIs outcomes. In contrast, wet pavement, low visibility, speeding, and overtaking increase the probabilities of surgical or fatal TBIs.

The study demonstrates that male patients had varied outcomes across observations as a random parameter. About 55.48% of male patients are more likely to experience surgical and fatal outcomes, while the remaining 44.52% are likely to receive non-surgical treatment, as indicated by the standard deviation value. Marginal effect analysis supports this, showing higher probabilities of both surgical and fatal outcomes for male patients. Male motorcycle users also have a greater likelihood of facing fatal consequences, as identified by heterogeneity in the mean. Another random parameter, 'rural road,' revealed that approximately 60.94% of crashes on rural roads are more likely to result in surgeries and fatalities, while 39.06% do not require surgery. Additionally, nighttime crashes on rural roads have a significantly higher probability of experiencing fatal outcomes.

The study also identified two key interaction effects that increase the risk of fatal TBIs. One interaction is the involvement of bus combined with low visibility conditions, which increases the probability of fatal TBIs. Another interaction is overtaking on wet pavements, which raises the chances of fatal TBIs. These combinations enhance the likelihood of high-impact, life-threatening TBIs.

Based on the findings of this study, it is essential to develop and implement targeted policies to reduce the incidence and impact of TBIs resulting from RTCs. Key recommendation includes enhancing pedestrian safety through the construction of signalize crosswalks, and adequate warning sign, especially in high-traffic areas. Enforcing speed limits and implementing speed control measures, particularly in rural and high-risk areas, can significantly reduce the incidence of crashes and TBIs. Creating dedicated lanes for buses and trucks can help manage traffic flow and reduce the likelihood of collisions and TBIs outcomes. Public awareness campaigns focused on the dangers of speeding, overtaking, the importance of helmet use for motorcycle users, and driving under poor visibility or wet conditions can a play a vital role in promoting safer driving behaviors. Investing in emergency response systems, particularly in rural areas, will ensure timely medical assistance, thereby reducing the fatality rates from TBIs. Police reports should thoroughly include all incidents, ensuring that every case is documented in detail. Additionally, identifying crash-prone areas and locations where secondary crashes are likely to occur is essential. To enhance the management and treatment of TBIs resulting from RTCs, it is essential to establish adequate casualty departments in both crash-prone and rural areas. It can significantly reduce the time it takes for patients to receive critical medical attention, thereby improving their chances of survival. Moreover, implementing an electronic medical records (EMR) system is necessary for gathering accurate data on patient care. This digital system would enhance the accuracy of the data collected, reduce the likelihood of errors, and ensure that patient records are easily accessible to healthcare providers. By adopting these policy measures, the goal of enhancing road safety and reducing the public health burden of TBIs can be effectively achieved.

## ACKNKOWLEDGEMNET

The authors thank the physicians and staff of Shaheed Ziaur Rahman Medical College Hospital, Bogura, for their support in conducting this study.

## AUTHOR CONTRIBUTIONS

The authors confirm contribution to the paper as follows: study conception and design: Azka Rodoshi Oishi, Md Jamil Ahsan, B M Tazbiul Hassan Anik; data collection and preparation: Azka Rodoshi Oishi; methodology, interpretation of results, manuscript preparation: Azka Rodoshi Oishi, Md Jamil Ahsan, B M Tazbiul Hassan Anik. All authors reviewed the results and approved the final version of the manuscript.